\documentclass[aps,twocolumn,groupedaddress]{revtex4}
\input epsf



\usepackage{graphicx}
\usepackage{dcolumn}

\begin{document}


\title{Dynamics of a financial market index after a crash}




\author{Fabrizio Lillo}
\email[]{lillo@lagash.dft.unipa.it}
\affiliation{Istituto Nazionale per la Fisica della Materia, Unit\`a di 
Palermo, Viale delle Scienze, I-90128, Palermo, Italy}



\author{Rosario N. Mantegna}
\email[]{mantegna@unipa.it}
\affiliation{Istituto Nazionale per la Fisica della Materia, Unit\`a di 
Palermo, Viale delle Scienze, I-90128, Palermo, Italy
and Dipartimento di Fisica e Tecnologie Relative, Universit\`a di Palermo,
Viale delle Scienze, I-90128, Palermo, Italy}



\begin{abstract}
We discuss the statistical properties of index returns in a 
financial market just after a major market crash. 
The observed non-stationary behavior of index returns is characterized
in terms of the exceedances over a given threshold. This characterization 
is analogous to the Omori law originally observed in geophysics. 
By performing numerical simulations and
theoretical modelling, we show that the nonlinear behavior observed
in real market crashes cannot be described by a GARCH(1,1) model.
We also show that the time evolution of the Value at Risk observed
just after a major crash is described by a power-law
function lacking a typical scale. 
\end{abstract}

\maketitle

\section{Introduction}
The time evolution of basic indicators of a financial
market as, for example, the volatility of a market index 
is often showing a non-stationary pattern \cite{Pagan}. 
In fact, tests
for whether volatility fluctuations of the size observed
in empirical time series could be due to estimation error 
strongly reject the hypothesis of constant volatility 
\cite{Schwert89}.

The volatility time series, both historical and implied,
experiences a dramatic increase at and immediately after
each financial crash \cite{Chen}. 
After the crash, the market volatility
shows a stochastic evolution characterized by a slow 
decay. 

A non-stationary time evolution of volatility directly 
implies a non-stationary time evolution of the returns of a
market index.
The purpose of this paper is to show that the dynamics 
of market index returns just after a crash presents a statistical 
regularity with respect to the number of times the
absolute value of index return exceeds a given threshold 
value.

The discovered statistical regularity is analogous to a 
statistical law discovered in geophysics more than a 
century ago and known today as the Omori law
\cite{Omori1894}.
The Omori law states that the number of aftershock
earthquakes
measured at time $t$ after the main earthquake decays with
$t$ as a power-law function. The observation of this specific 
functional form in the stochastic relaxation of the market 
statistical indicators to its typical values implies that the 
relaxation dynamics of a market index just after
a financial crash is not characterized by a typical scale.
The lack of a typical relaxation scale 
has direct implication on the estimation of risk measures as,
for example, the Value at Risk of a stock portfolio \cite{Duffie97}.

The structure of the paper is as follows. In Section 2
we report on the empirical properties of aftercrash 
market index time series. The discussion concerns the 
1-minute Standard and Poor's 500 index returns 
recorded after the Black Monday financial crash. 
Section 3 shows that the GARCH(1,1) model is unable to 
describe the empirical properties of the previous section. 
Section 4 discusses the estimation of the Value at Risk 
during the time period immediately after a financial crash. 
The paper ends with some concluding remarks.

\section{Empirical properties of aftercrash sequences}
In this section, we characterize the time series of 
ultra high frequency returns after a major market crash.
The variable investigated is the one-minute logarithm 
changes of a financial index $r(t)$.
A direct characterization of the time evolution of this random 
process is extremely difficult in the time period after a market
crash. This is due to the fact that the aftercrash
period is highly non-stationary because the financial market
needs some time to be back to a "normal" period. 
In order to characterize the aftercrash return time 
series we make use of a statistically robust method. 
Specifically, we quantitatively characterize the time 
series of index returns by investigating the number
of times $|r(t)|$ is exceeding a given threshold value 
\cite{lillo2002} in the non-stationary time period. 

A similar approach is used in the investigation of the number $n(t)$
per unit time of aftershock earthquakes above a given threshold 
measured at time $t$ after the main 
earthquake. This quantity is well described in geophysics by the Omori 
law \cite{Omori1894}. The Omori law $n(t) \propto t^{-p}$ says 
that the number of aftershock earthquakes per unit time measured at
time $t$ after the main earthquake decays as a power law.
In order to avoid divergence at $t=0$, Omori law is often
rewritten as
\begin{equation}
n(t)=K(t+\tau)^{-p},
\end{equation}
where $K$ and $\tau$ are two positive constants.
An equivalent formulation of the Omori law, which is more suitable for
comparison with real data, can be obtained by integrating
Eq. (1) between $0$ and $t$. In this way the cumulative 
number of aftershocks after the main 
earthquake observed until time $t$ is
\begin{equation}
N(t)=K[(t+\tau)^{1-p}-\tau^{1-p}]/(1-p),
\end{equation}
when $p \ne 1$ and $N(t)=K\ln(t/\tau+1)$
for $p=1$. 

When the process is stationary the frequency of aftershock $n(t)$
is on average constant in time and therefore the 
cumulative number $N(t)$ increases linearly in time. 
We have tested that $N(t)$ increases approximately 
linearly in a market period of approximately
constant volatility such as, for example, the 1984 year.
For independent identically distributed random time series
it is possible to characterize $n(t)$ in terms of an 
homogeneous Poisson process \cite{Embrechts}.

As an example of the behavior of $N(t)$ empirically observed 
after a market crash,
we investigate the index returns during the time period just
after the Black Monday crash (19 October 1987)
occurred at New York Stock Exchange 
(NYSE). This crash was one of the worst crashes occurred in the entire 
history of NYSE. The Standard and Poor's 500 Index (S\&P500)
went down $20.4\%$
that day. In our investigation, we select a 60 trading day aftercrash
time period ranging from 20 October 1987 to 14 January 1988.
For the selected time period, we investigate the one-minute return time
series of the S\&P500 Index. The 
unconditional one-minute volatility is equal to 
$\sigma=4.91 \times 10^{-4}$.
In Fig. 1 we show the cumulative number of events $N(t)$
detected by considering all the occurrences observed when the the 
absolute value of index return exceeds a threshold value $\ell=4 \sigma$. 
We observe a nonlinear behavior in the entire period. Fig. 1 also shows 
our best nonlinear fit performed with the functional form of Eq. (2).
The agreement between empirical data and the functional form of the Omori 
law is pretty good. 
With a single time series it is not possible to have a good description 
of $n(t)$ due to the discreteness of the data. In order to
solve this problem we compute $n_{av}(t)$ which is
the running average of $n(t)$ in a 
sliding window of $200$ trading minutes.
In the inset of Fig. 1 we show this quantity as a function of time 
in a bilogarithmic plot. A power-law behavior is observed for almost
two decades and the exponent of the best power-law 
fit is in agreement with the exponent $p$ obtained by fitting $N(t)$. 

We observe a similar behavior when we set the threshold value
$\ell$ equal to $5 \sigma$, $6 \sigma$ and $7 \sigma$.
The best fit of the exponent $p$ slightly increases when  $\ell$ 
increases and eventually converges to a constant value.

This paradigmatic behavior is 
not specific of the Black Monday 
crash of the S\&P 500 index. In fact, we observe 
similar results also for a 
stock price index weighted by market 
capitalization for the time periods 
occurring after the 27 October 1997 and the 
31 August 1998 stock market crashes \cite{lillo2002}. 
This index has been computed  
selecting the 30 most capitalized stocks 
traded in the NYSE and by using the high-frequency
data of the {\it Trade and Quote} (TAQ) database 
issued by the NYSE.

\begin{figure}[tc]
\epsfxsize=3.2in
\epsfbox{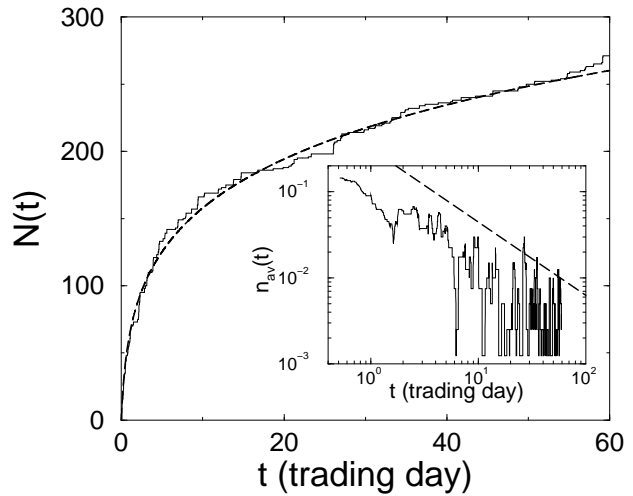}
\caption{Cumulative number $N(t)$ of the number of times $|r(t)|$ is
exceeding the threshold $\ell=4\sigma$ during the 60 
trading days immediately after the Black Monday financial crash.
The parameter $\sigma$ is the standard
deviation of the process $r(t)$ computed over the entire investigated period.
The dashed line is the best fit of Eq. (2). The value of the exponent $p$
is  0.85. In the inset, we show $n_{av}(t)$ 
which is a moving average of $n(t)$ in a sliding
window of 200 trading minutes. The dashed line is a power-law curve with
an exponent equal to $-0.85$ and is shown as a guide for the eye}
\label{fig1}
\end{figure}
 
Our empirical results show that 
index return cannot be modeled in terms of independent 
identically distributed random process after a big market crash.
In ref. \cite{lillo2002} we introduce a simple stochastic model 
which is able to explain the Omori law in financial time series. 
Specifically we assume that the stochastic variable $r(t)$ 
is the product of a deterministic time dependent 
scale $\gamma(t)$ times a stationary stochastic process $r_s(t)$
during the time period after a big crash.
Under these assumptions, the number of events 
of $|r(t)|$ larger than $\ell$ observed at time $t$ is proportional to
\begin{equation} 
n(t) \propto 1-F_s\left(\frac{\ell}{\gamma(t)}\right),
\end{equation} 
where $F_s(r_s)$ is the cumulative
distribution function of the random variable 
$r_s(t)$.

By assuming that the stationary return probability density function
behaves asymptotically as a power-law 
\begin{equation} 
f_s(r_s)\sim \frac{1}{r_s^{\alpha+1}},
\end{equation}
and that the scale of the stochastic process
decays as a power-law
$\gamma(t)\sim t^{-\beta}$, the number of events above threshold is
power-law decaying as $n(t)\sim (\gamma(t)/\ell)^{\alpha}\sim1/t^p$. 
The exponent $p$ is given by 
\begin{equation} 
p=\alpha~\beta.
\end{equation}
The previous relation links the exponent $p$ governing the
number of events exceeding a given threshold to the
$\alpha$ exponent of the stationary power-law return cumulative distribution
and to the $\beta$ exponent of the power-law decaying
scale. 

\begin{figure}[tc]
\epsfxsize=3.2in
\epsfbox{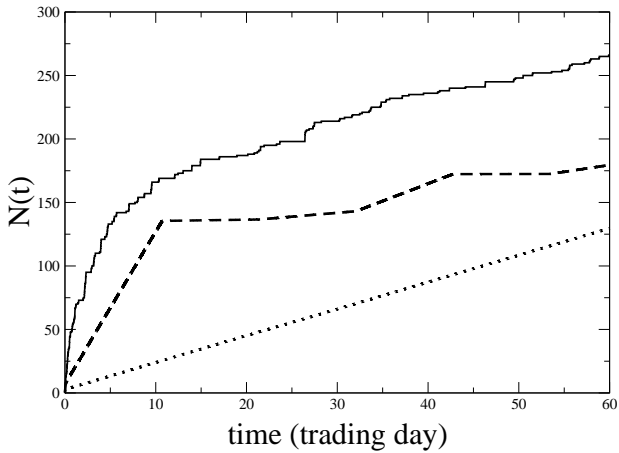}
\caption{The dotted line is the mean value of the cumulative number $N(t)$ of 
the number of times $|r(t)|$ is exceeding the threshold $\ell=4\sigma$ 
over $10^4$ simulations obtained with a GARCH(1,1) model with parameters 
estimated in a period of 60 trading days immediately after the 
Black Monday financial crash. The dashed line is the mean of 
$N(t)$ over $10^4$ simulations obtained with a GARCH(1,1) 
model with parameters estimated in 6 nonoverlapping windows of 
10 days each. As a reference we show also $N(t)$ for the real data
already shown in Fig. 1 (solid line).}
\label{fig2}
\end{figure}

The hypotheses of our model
are consistent with recent empirical results. In fact, a return probability
density function
characterized by a power-law asymptotic behavior has been
observed in the price dynamics of several stocks \cite{Lux96,Gopi98} and
a power-law or power-law log-periodic decay of implied volatility has 
been observed in the S\&P500 after the 1987 financial
crash \cite{Sornette96}. 
In \cite{lillo2002} we have tested the hypotheses of our model by 
measuring independently the exponents $\alpha$ and $\beta$
from real data and we have shown that the relation between exponents
described by Eq. (5) is satisfied in all the three investigated crashes.


\section{Comparison with GARCH(1,1) model}

Here we compare the empirical behavior of a market index after a 
major crash with the predictions of a
simple autoregressive model, specifically the GARCH(1,1) model
\cite{Bollerslev86} described by the equation
\begin{equation}
\sigma^2_t=\alpha_0+\alpha_1r^2_{t-1}+\beta_1 \sigma^2_{t-1},
\end{equation}
where $r_t$ is price return and $\sigma_t$ is the volatility.  
To this end we estimate the best value of the parameters of the 
GARCH(1,1) model. 
The estimation has been performed with the G@RCH 2.3 package 
\cite{garch}. The best estimation of the parameters on the time series
of one minute logarithmic price change during 60 trading days after 
Black Monday crash are $\alpha_0=2.87~10^{-8}$, $\alpha_1=0.38$ and 
$\beta_1=0.54$. With these parameters
we generate $10^4$ surrogate time series according to Eq. (6) and we
compute the average behavior of $N(t)$. 
To mimic the dynamics after the crash, we set as 
an initial condition a large value of return in each realization.
Fig. 2 shows $N(t)$ versus
time for real data and GARCH(1,1) model (dotted line) with a 
threshold $\ell=4\sigma$ (the same as in Fig. 1). 
The GARCH(1,1) time series converges to its stationary phase very quickly
and it is unable to show a significant nonlinear behavior.

In order to take into account the non-stationary behavior of the return
time series after a crash we have performed a different analysis.
Specifically we divide the 60 trading day period in 6 nonoverlapping
time intervals of 10 trading days and we estimate the GARCH(1,1) parameters
specific for each interval. We then generate $10^4$ GARCH(1,1) surrogate time 
series using the estimated parameters for each subinterval. 
The mean value of $N(t)$ as a function of time obtained with this procedure
is also shown in Fig. 2 as a dashed line.
In each subinterval $N(t)$ increases linearly for almost all the period.
This is due to the fact that the non-stationary part is extremely small
compared with the time scale of the figure.

Below we theoretically describe the behavior of $N(t)$ expected for a 
GARCH(1,1) time series. First of all we note that 
the expectation value of $\sigma^2_t$ conditioned to the value $\sigma^2_0$
is given by \cite{Baillie92}
\begin{equation}
E(\sigma^2_t|\sigma^2_0)=\sigma^2_0(\alpha_1+\beta_1)^t+
\alpha_0\frac{1-(\alpha_1+\beta_1)^t}{1-(\alpha_1+\beta_1)}.
\end{equation}
This equation shows that, under the assumption of finite unconditional 
variance $\alpha_1+\beta_1<1$, the mean value of the scale of the process 
decays exponentially to the unconditional value 
in a GARCH(1,1) time series. The decaying time is
given by $\tau=-1/\ln(\alpha_1+\beta_1)$. 
With the parameters estimated we have $\tau=12$ trading minutes
in the case considered in Fig. 2. 

\begin{figure}[tc]
\epsfxsize=3.2in
\epsfbox{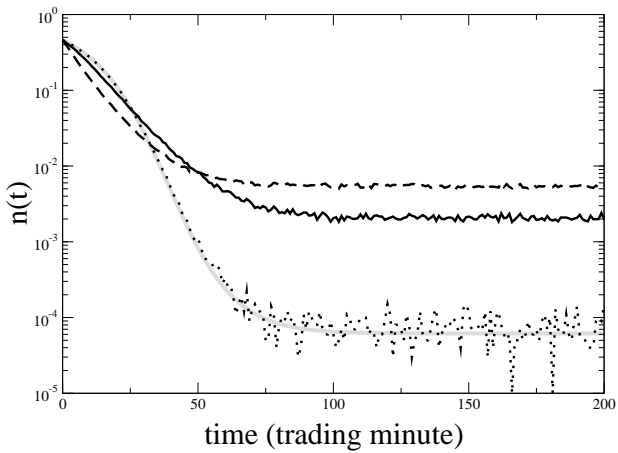}
\caption{Mean of $n(t)$, the number per unit of time
that $|r(t)|$ is exceeding the threshold $\ell=4\sigma$, for 
$10^5$ numerical simulation of three
different GARCH(1,1) models with the same value of $\alpha_0$ and of 
$\alpha_1+\beta_1=0.92$.
The values of $\alpha_1$ are $0.02$ (dotted line),
$0.18$ (solid line) and $0.38$ (dashed line).
This last value is the one estimated from the real 
time series of Black Monday aftercrash period.
The gray line is the prediction based on the approximation
of Gaussian density.}
\label{fig3}
\end{figure}

Since the variance of the GARCH(1,1) model depends only on $\alpha_0$ 
and on $\alpha_1+\beta_1$, we perform a set of numerical simulations
of GARCH(1,1) time series with the same value of $\alpha_0$ and by
keeping constant $\alpha_1+\beta_1=0.92$, which are the values
observed in the real data. 
Different time series of simulations are obtained by changing
the value
of $\alpha_1$ which controls the leptokurtosis of the 
unconditional density.
Fig. 3 shows the average of $n(t)$ over $10^5$ realizations of
a GARCH(1,1) model with $\alpha_1=0.02$, $\alpha_1=0.18$ and 
$\alpha_1=0.38$. 
This last value corresponds to the one estimated from the real data.
In the first two cases the unconditional kurtosis is finite,
whereas for $\alpha_1=0.38$ the fourth moment diverges.
The value of the threshold is set to $\ell=4\sigma$ in all cases.
The unconditional value of $n(t)$ is larger for larger values
of $\alpha_1$. This is expected because the unconditional densities
have the same variance and the tail are fatter for larger values
of $\alpha_1$.
For all values of $\alpha_1$, $n(t)$ decays to the stationary 
value in less than a trading day. The decay is essentially
exponential and a best fit with a function 
$n(t)=a+b~e^{-ct}$ gives a good agreement.

Fig. 3 also shows that the larger is $\alpha_1$ (and therefore 
the unconditional kurtosis), the faster is the decay of $n(t)$
for very small values of $t$. 
This empirical behavior can be modelled and hereafter we show
how it is possible to find an estimate of $n(t)$ for
small values of $t$.
Let us consider that at time $t=0$
the crash occurs and the return is $r_0$. At time $t=1$ the 
density is Gaussian with zero mean and variance given by 
$\sigma_1=\alpha_0+(\alpha_1+\beta_1)r_0^2$.
However at subsequent times the density is no more Gaussian. 
As a first approximation we can assume that the density is Gaussian
with time-dependent variance described by (7). This is a very rough
approximation, which is valid only when $\alpha_1$ is very small. The gray
line in Fig. 3 is $n_{gau}(t)=erfc(\ell/\sqrt{2\sigma^2_t})$
which is the function obtained by using this approximation.
For large values of $t$, $n(t)$ tends to a constant value as $e^{-t/2\tau}$.
The agreement is very good in the case $\alpha_1=0.02$. However, 
for larger values of $\alpha_1$ a better approximation is needed to 
explain the results of numerical simulations. 
A better approximation can be obtained by taking into account higher moments.
Even if at time $t=1$ the density is Gaussian,
at $t=2$ the density is no more Gaussian. 
Nevertheless we can find the
exact expression for the expected value of the first four moments
at $t=2$ conditioned to the information at time $t=0$.
One finds
\begin{eqnarray}
E_0(r_2)=E_0(r_2^3)=0, \\
E_0(r_2^2)=\alpha_0+(\alpha_1+\beta_1)\sigma_1^2\equiv s_2^2,\\
E_0(r_2^4)=6\alpha_1^2\sigma_1^4+E_0(r_2^2).
\end{eqnarray}
Under the assumption that the density is not very different 
from Gaussian at time $t=2$, we can approximate the density 
with a first order Edgeworth expansion \cite{Feller}
\begin{equation}
p(r_2)=\frac{e^{-r_2^2/2s_2^2}}{\sqrt{2\pi}}
\left(1+He_4\left(\frac{r_2}{s_2}\right)\frac{\kappa_4(t=2)}{4!~s_2^4}
\right),
\label{edgeworht}
\end{equation}
where $He_4(x)$ is the fourth Chebyshev-Hermite polynomial and
$\kappa_4(t)$ is the fourth cumulant of $r_t$.
By integrating $p(x)$ in Eq. (\ref{edgeworht}) between
$\ell$ and infinity we can find an expression for
$n(t=2)$. By performing the integration,
we observe two different regimes: (i) when 
$\ell^2<3(\alpha_0(1+\alpha_1+\beta_1)+(\alpha_1+\beta_1)^2r_0^2)$,
$n(t=2)$ decreases when $\alpha_1$ increases; (ii) when 
$\ell^2>3(\alpha_0(1+\alpha_1+\beta_1)+(\alpha_1+\beta_1)^2r_0^2)$,
$n(t=2)$ increases when $\alpha_1$ increases.
Table 1 shows two numerical examples corresponding to the two regimes.
The results of Table 1 show that the agreement between simulated values  
and predicted values of $n(t)$
is quite good especially for small values of $\alpha_1$. The fact that
for large values of $\alpha_1$ the agreement is less precise is
due to the significant contribution coming from the moments higher 
than the fourth that have been neglected in the Edgeworth expansion
of Eq. (\ref{edgeworht}). 
One could improve the forecast of $n(t)$ by extending the Edgeworth
expansion to higher order and by using the approximate density so
obtained to calculate a better estimation of $n(t)$.

The technique of Edgeworth expansion can also be used to forecast 
the value of $n(t)$ for value of
$t$ larger than $2$. Our numerical simulations show that for moderate
values of $\alpha_1$ and/or short forecast horizons the predictions
work quite well. For larger value of $\alpha_1$ (especially when
the fourth moment of the unconditional density is infinite)
the prediction is valid only for small values of $t$.

These theoretical and numerical observations confirm that the GARCH(1,1)
is unable to model the power law decay of $n(t)$ observed in real time
series after a major market crash.

\begin{table}
\caption{Simulated and theoretical values of $n(t=2)$ 
in a GARCH(1,1,) model for different values
of $\alpha_1$ and for the two regimes described in the text. 
Case (i) corresponds to $r_0=3.4~10^{-3}$
and case (ii) corresponds to $r_0=1~10^{-3}$. The other parameters are:
$\alpha_0=2.87~10^{-8}$, $\ell=4\sigma=2.4~10^{-3}$.} 
\begin{tabular}{cc|cc|cc}
 &&\multicolumn{2}{c}{case (i)}&\multicolumn{2}{c}{case (ii)}\\
 $\alpha_1$& $\beta_1$ & $n_{sim}(2)$ & $n_{teo}(2)$ & $n_{sim}(2)$ & $n_{teo}(2)$ \\ \hline
0.02 & 0.90 & $4.41~10^{-1}$ &  $4.41~10^{-1}$ & $1.11~10^{-2}$  & $1.15~10^{-2}$\\
0.18 & 0.74 & $4.34~10^{-1}$ &  $4.31~10^{-1}$ & $1.26~10^{-2}$  & $1.40~10^{-2}$\\
0.38 & 0.54 & $4.11~10^{-1}$ &  $3.95~10^{-1}$ & $1.75~10^{-2}$  & $2.26~10^{-2}$ \\
0.58 & 0.34 & $3.73~10^{-1}$ &  $3.33~10^{-1}$ & $2.27~10^{-2}$  & $3.75~10^{-2}$ \\
\end{tabular}
\end{table}

\section{Value at Risk after a crash}

The empirical evidence and the theoretical model 
shown in the previous sections have direct relevance for risk 
management. One of the most widely used measure of risk is the
so called Value at Risk (VaR) \cite{Duffie97}. The VaR is the 
most probable worst
expected loss at a given level of confidence over a given time horizon.
Here we indicate the probability density for index return
$r$ over the time horizon $\tau$ as $f_{\tau}(r)$. The VaR 
$\Lambda_{VaR}$ associated to a certain probability of loss
$P_{VaR}$ is defined implicitly by 
\begin{equation}
\int_{-\infty}^{-\Lambda_{VaR}} f_{\tau}(r)~dr=P_{VaR}.
\end{equation}

In stationary market conditions, the VaR is determined by the 
the density $f_{\tau}(r)$. In Section 2 we have discussed a simple
market model able to describe quantitatively the non-stationary
evolution of index return time series after a market crash. In this model
the shape of the return density remains constant but its scale 
changes in time. By using the relation $r(t)=\gamma(t)r_s(t)$,
it is direct to show that the (instantaneous) VaR after a market crash is not 
constant but changes in time as  
\begin{equation}
\Lambda_{VaR}(t)=\gamma(t)\Lambda^{(s)}_{VaR},
\end{equation}
where $\Lambda^{(s)}_{VaR}$ is the constant VaR obtained by using
the stationary probability density function $f_s(r_s)$. We have seen 
\cite{lillo2002} that the occurrence of the
Omori law after a market crash is consistent with a time evolution
of the scale $\gamma(t)=c_1 t^{-\beta} +c_2$. Eq. (13) indicates
that after a big market crash the VaR decreases in time as a power-law
with exponent $\beta$ (close to $0.2 \div 0.3$ in the investigated crashes)
and eventually relaxes to a constant value.   
\section{Conclusions}
Our empirical observations show that the statistical properties 
of index return time series after a major financial crash are essentially
different from the ones observed far from the crash.
Other examples of statistical properties of market 
which are specific of the aftercrash period have been observed in 
the investigation of cross-sectional quantities 
computed for a set of stocks before, at and after 
financial crashes \cite{Lillo2000,Lillo2001}.
The time period just after a crash is characterized by a relaxation
to the typical market phase that can be modelled in terms of a
power-law decay of the typical scale of index returns.
We have shown that this observation rules out the possibility
that the statistical properties of the time evolution of index 
returns can be efficiently modelled in terms of simple autoregressive
models, such as the GARCH(1,1) model, in the non-stationary period
after a crash.
Our modelling of the aftercrash period also suggests that the Value
at Risk of a financial portfolio measured just after a financial
crash evolves in a non trivial way characterized by a power-law 
evolution lacking a typical scale. 
In summary we conclude that the time period after a major market
crash is characterized by statistical regularities which are
specific to such a time period and not well described by models 
which presents a typical time scale in their stochastic evolution.
The authors thank INFM ASI and MURST for financial support.



\end{document}